# On the thermodynamic analogy of intracellular diffusivity fluctuations

**Yuichi Itto**[1,2]

[1] Science Division, Center for General Education, Aichi Institute of Technology, Aichi 470-0392, Japan

[2] ICP, Universität Stuttgart, 70569 Stuttgart, Germany

**Abstract.** Two recent topics on a formal thermodynamic analogy of intracellular diffusivity fluctuations observed experimentally in normal/anomalous diffusion are reported. Not only the analogs of the quantity of heat and work as well as the internal energy but also that of the Clausius inequality are identified. Then, the analog of the heat engine is constructed to characterize extraction of the diffusivity change as the analog of work during a cycle, the efficiency of which is formally equivalent to that of the Carnot engine, making the total change of the entropy concerning the fluctuations vanish. The effect of the slow variation of the fluctuations on the efficiency is also briefly discussed.

## 1. Introduction

A recent experimental breakthrough in the physics of diffusion in living cells would be the development of the technique of single-particle tracking [1-4], which has revealed a variety of exotic physical phenomena at the level of individual trajectory. One such example is heterogeneous diffusion with diffusivity fluctuations: over the entire region of the living cell, the diffusion coefficient of (macro)molecules slowly fluctuates locally, the statistical form of which obeys several definite laws (see, e.g., Refs. [5,6] for recent reviews).

Let $\overline{\Delta^2}$ be the time-averaged mean square displacement [3,5,6], i.e., the average of square displacements over a certain duration of time in a given individual trajectory, that scales for elapsed time, $\tau$, as $\overline{\Delta^2} \sim D\tau^{\alpha}$, where $D$ is the diffusivity, i.e., the diffusion coefficient and $\alpha$ is the diffusion exponent taking a certain positive value. Normal diffusion reads $\alpha = 1$, whereas the case of $\alpha \neq 1$ is referred to as anomalous diffusion [7] widely studied in the literature. It is noticed that the dimension of $D$ depends on $\alpha$. Among the above-mentioned definite laws for the diffusivity fluctuation distribution, the exponential law, which keeps an outstanding position in experiments, is of central importance:

$$P_0(D) \sim e^{-\frac{D}{D_0}} \tag{1}$$

with $D_0$ being the average value of $D \in [0, \infty)$, (where the subscript 0 indicates a reference distribution). Several examples in the case of normal diffusion are lipid vesicles in a solution of F-actin filaments [8], transmembrane proteins and lipid molecules on cell membrane [9,10], whereas RNA-protein particles in bacteria or yeasts [11,12] are the case for anomalous diffusion. The statistical fluctuation distribution is known to be crucial for describing the displacement distributions (see, e.g., Refs. [13-15]), which typically exhibit non-Gaussian behavior. Such a description is given by the superposition of the local stochastic process with the fluctuating diffusivity in terms of the fluctuation distribution, showing the relevance of the statistical property of the displacements to that of the fluctuations.

Nowadays, there is growing interest in understanding the effect of environmental conditions on the diffusivity, the change of which plays an essential role, for example, for tuning the rates of biochemical reactions in cells, as seen, e.g., in Refs. [16-18]. In this respect, several experimental facts are mentioned. In the mechanical effect, at the statistical level, the diffusivity decreases (increases) under compression [19-22] (stretch [20], i.e., expansion) of cells, whereas almost no change is observed for the diffusion exponent [22], while cell volume/shape is restored by relaxing compression [19,22]. Then, from the relevance in the above, the stability of the exponential law in Eq. (1) under such a compression/expansion is expected since the statistical property of displacements is robust [20,22]. On the other hand, the thermal effect is naturally considered to be relevant under the assumption that like in the Einstein relation [23], the diffusivity is proportional to local temperatures in cells, which has in fact been supported, e.g., in Refs. [15, 24-26] (see also the discussion in Sec. 2). Indeed, local temperatures have experimentally been measured in living cells, see, e.g., Refs. [27-30]. It is not too much to say that problems of controlling the change of the diffusivity by adjusting the environmental effect are highly challenging (see, e.g., Ref. [31] and references therein toward such a direction). The discussions we will develop here are expected to be useful for analyzing/extracting such a change in a thermodynamic-like formalism.

A possibility is therefore open for investigating fundamental properties of diffusivity fluctuations in terms of the environmental effect. Taking into account the above facts, the change of the diffusivity with the form of the exponential law being kept fixed is of particular interest.

In this article, we report two recent topics on a formal thermodynamic analogy of intracellular diffusivity fluctuations [32,33]: one is associated with the analogs of the thermodynamic laws, and the other is about that of the heat engine. Regarding the diffusivity as the analog of the system energy, the analogs of the internal energy, work, and the quantity of heat are identified. The Clausius-like inequality is also established for the entropy concerning the fluctuations. The analog of the heat engine is then constructed to quantify how much the diffusivity change as the analog of work is extracted during a cycle, the processes of which are realized by compression/expansion of cells and the change of temperature. Its efficiency formally has that of the Carnot engine, making the total entropy change in the cycle vanish. We also make a comment on the effect of slowly varying fluctuation on the efficiency by examining the dynamical behavior of the average diffusivity, which is not discussed in the work in Ref. [33].

## 2. Analogs of the thermodynamic laws

Let us regard a cell as a medium consisting of many small local regions or "blocks" with a set of different diffusivities $\{D_i\}_i$, which obeys some statistical fluctuation distribution, $P(D_i)$, and is denoted purposely in the discrete case, for normal or anomalous diffusion. This distribution can slightly deviate from the exponential law in Eq. (1), in general, since diffusivity fluctuations are supposed to slowly vary, the time scale of which is much larger than that of a typical local dynamics of a random walker. In other words, the medium is quasistationary, i.e., nonstationary on a large time scale. As mentioned in the Introduction, a basic assumption is that the diffusivity is proportional to temperature in a given local block: $D_i = cT_i$, where $c$ and $T_i$ are a proportionality factor characterizing mobility and the $i$ th value of local temperature, respectively. Regarding this point, we mention the following in the case of anomalous diffusion, in particular. The works in Refs. [25,26] have shown the linear dependence of temperature on the (conventional) mean square displacement for large elapsed time, which turns out to be equivalent to the time-averaged mean square displacement (see Ref. [3]) since a generalized Langevin equation discussed there for characterizing viscoelastic nature of cells describes the process of fractional Brownian motion [34] with the same diffusion property (see Ref. [24]). It is noted that $c$ does not vary drastically over the local blocks since the diffusion exponent is (approximately) constant, here.

We express the exponential law in Eq. (1) as follows:

$$P_0(D_i) = \frac{1}{Z} e^{-\frac{D_i}{cT}} \tag{2}$$

with $Z = \sum_i e^{-D_i/(cT)}$, where $D_0 = cT$ is used with $T$ being the average value of local temperatures in the continuum limit. This allows us to see that at the statistical level, the exponential law formally takes the canonical distribution in Boltzmann-Gibbs statistical mechanics if the diffusivity is regarded as the analog of the system energy. Accordingly, the medium is in the "equilibrium state" in the case when $P(D_i) = P_0(D_i)$. It is noticed that the distribution in Eq. (2) actually does not depend on $c$. From these, the average diffusivity is identified with the analog of the internal energy, (which is justified from a relation as seen below):

$$U_D = \sum_i D_i \, P(D_i). \tag{3}$$

We shall consider a thermodynamic-like process, along which the average diffusivity can change on the time scale much larger than that of variation of diffusivity fluctuations. There, any two states of the medium are distinguished, in which the local properties of diffusivity fluctuations are infinitesimally different from each other. Therefore, the change of the average diffusivity along this process is given by

$$\delta U_D = \sum_i D_i \, \delta P(D_i) + \sum_i P(D_i) \, \delta D_i, \tag{4}$$

where $\delta P(D_i)$ stands for the change of the statistical form of the fluctuation distribution and $\delta D_i$ describes the change of the diffusivity. Taking into account Eq. (3), the analogs of work and the quantity of heat are identified with

$$\delta' W_D = -\sum_i P(D_i) \, \delta D_i \tag{5}$$

and

$$\delta' Q_D = \sum_i D_i \, \delta P(D_i), \tag{6}$$

respectively. Thus, we obtain the following analog of the first law of thermodynamics [35]:

$$\delta' Q_D = \delta U_D + \delta' W_D. \tag{7}$$

Now, we see how the changes on the right-hand side of Eq. (4) are realized in the case of the exponential law. Since the distribution in Eq. (2) is independent of $c$, $\delta P(D_i)$ originates from the change of $T$ while the value of the diffusivity is kept unchanged, whereas $\delta D_i$ comes from the change of $c$ while the form of the fluctuation distribution is kept fixed. In particular, from the fact that the diffusivity decreases (increases) under compression (expansion) of cells, the latter implies that $c$ plays a role analogous to external parameter. Therefore, the analogs in Eqs. (5) and (6) under Eq. (2) are found to be given by

$$\delta' W_D = -\frac{\partial \langle D_i \rangle}{\partial c} \delta c \tag{8}$$

and

$$\delta' Q_D = \frac{\left\langle \left( D_i - \langle D_i \rangle \right)^2 \right\rangle}{c T^2} \delta T, \tag{9}$$

where $\delta c$ and $\delta T$ stand for the changes of $c$ and $T$, respectively, and $\langle A_i \rangle \equiv \sum_i A_i P_0(D_i)$. Equivalently, they are expressed with $Z$ as

$$\delta' W_D = -T^2 \frac{\partial \ln Z}{\partial T} \delta c \tag{10}$$

and

$$\delta' Q_D = c \frac{\partial}{\partial T} \left( T^2 \frac{\partial \ln Z}{\partial T} \right) \delta T. \tag{11}$$

Subsequently, we achieve the analog of the second law of thermodynamics for the entropy concerning diffusivity fluctuations, i.e., a measure of uncertainty about the local property of the fluctuations. As shown in Refs. [32,33] (see also Refs. [36,37]), in a way similar to the one for deriving the Shannon entropy [38], such an entropy is introduced based on the local blocks independent of each other with respect to the diffusivity because of individual trajectories: $S[P] = -\sum_i P(D_i) \ln P(D_i)$. Then, the existence of the large time-scale separation for slow variation of the fluctuations may allow us to use the maximum entropy principle (see Refs. [39,40] for a recent development along this line) for $S$ with the constraints on the normalization of the fluctuation distribution and the expectation value of the diffusivity, the associated Lagrange multipliers with which are chosen in an appropriate manner, giving rise to the exponential law in Eq. (1) (in the discrete case). As we shall see below, what is essential here is to employ the following entropy:

$$\tilde{S}[P] = c\, S[P], \tag{12}$$

for which it is understood that the maximum entropy principle with $\tilde{S}$ also leads to the exponential diffusivity fluctuation under the redefinition of the Lagrange multipliers.

Keeping this in mind, we evaluate how the entropy change behaves in the case when $P(D_i)$ is different from $P_0(D_i)$, (which is based on a basic observation in Ref. [41]). Let us quantify the difference between them by the use of the following Kullback-Leibler relative entropy [42]: $K[P \,\|\, P_0] = \sum_i P(D_i) \ln\left[\, P(D_i) / P_0(D_i) \,\right]$, which is positive semidefinite and vanishes if and only if $P(D_i) = P_0(D_i)$. Then, we shall write the change of the fluctuation distribution as $\delta P(D_i) = \left\{\, \gamma P^*(D_i) + (1-\gamma) P(D_i) \,\right\} - P(D_i)$, where $P^*(D_i)$ is some fluctuation distribution fulfilling $K[P^* \,\|\, P_0] \le K[P \,\|\, P_0]$ and $\gamma$ is a constant in the range $0 < \gamma < 1$. With this, the change of the relative entropy is given by $\delta_P K[P \,\|\, P_0] = K[\,\gamma P^* + (1-\gamma) P \,\|\, P_0\,] - K[\, P \,\|\, P_0\,]$, where $\delta_P$ describes the change in terms of $P(D_i)$, and turns out not to be positive: $\delta_P K[P \,\|\, P_0] \le \gamma \left\{\, K[P^* \,\|\, P_0] - K[P \,\|\, P_0] \,\right\} \le 0$, in which the convexity of the relative entropy, $K[\,\gamma P^* + (1-\gamma) P \,\|\, P_0\,] \le \gamma\, K[P^* \,\|\, P_0\,] + (1-\gamma)\, K[P \,\|\, P_0\,]$, has been used. Under the normalization of $P(D_i)$, the change itself is calculated to be $\delta_P K[P \,\|\, P_0] = -\,\delta S[P] + \delta' Q_D / D_0$. Thus, in the case when $c$ is fixed, we arrive at the following analog of the Clausius inequality:

$$\delta \tilde{S}[P] \ge \frac{\delta' Q_D}{T}, \tag{13}$$

while the equality $\delta\tilde{S}[P_0] = \delta' Q_D / T$ comes from $\delta_P K[P \| P_0] = 0,$ which holds for $P(D_i) = P_0(D_i).$ This fact shows that $\tilde{S}$ in Eq. (12) is analogous to the thermodynamic entropy, together with justification of Eqs. (5) and (6).

In addition, from the above observations, we see how the identification in Eq. (3) is appropriate. In the continuum limit, the derivative of the entropy $\tilde{S}$ in Eq. (12) with the exponential law in Eq. (2), in which the entropy calculated as $S = 1 + \ln D_0$ is determined up to an additive constant from the dimensional reason of $P_0(D),$ with respect to the average diffusivity $D_0$ immediately gives the following relation in the case when $C$ is fixed [32,33,36]: $\partial\tilde{S} / \partial D_0 = 1/T.$ At the statistical level, this is analogous to the thermodynamic relation concerning temperature if $D_0$ is identified with the internal energy.

Closing this section, we shall point out the following. A relevance of a stretched exponential law of diffusivity fluctuations to a stretched exponential form of displacement distributions, which is studied in Ref. [43] for diffusion of phospholipids and proteins in crowded environments, has been discussed in Ref. [14] in the description mentioned in the Introduction. In a recent work in Ref. [44], where statistical mechanics for systems with unstable interactions has been developed, a stretched exponential distribution has been introduced as a generalized canonical distribution for describing equilibrium states of such systems, which depends on a single exponent characterized by a system Hamiltonian and becomes reduced to the ordinary canonical distribution for systems with stable interactions with the exponent being unity, whereas the conventional thermodynamic relations hold. Accordingly, the present formal analogy is expected to be extended in such a way that all the relevant quantities have the dependence of the exponent of the stretched-exponential diffusivity fluctuation, which should give a physical insight into its robustness in terms of a structural similarity with thermodynamics.

**3. Analog of the heat engine**

Next, we develop the analog of the heat engine for the exponential diffusivity fluctuation in Eq. (2) in the thermodynamic-like processes and examine its efficiency in order to characterize how much the diffusivity change as the analog of work in Eq. (8) can be extracted during a cycle constructed by the processes. This is the change of experimental relevance, since the stability of the exponential law under the process of compression/expansion of cells is expected as mentioned in the Introduction and temperature is supposed to remain constant in such a process.

Analogously to thermodynamics, let us define the analog of pressure by $p_D = -\partial \langle D_i \rangle / \partial c,$ which is calculated to be $p_D = -T$ in the continuum limit. Due to the analogy of $c$ to external parameter, we interpret the effect of compression (expansion) as the decrease (increase) of the value of $c$ in $D_0 = cT.$ From Eq. (7), it is seen that when $c$ decreases (increases), $\delta U_D$ becomes negative (positive), given a constant value of $T,$ i.e., $\delta' Q_D = 0,$ and $\delta' Q_D$ becomes positive (negative) while $\delta U_D = 0.$ Thus, we consider a cycle $\mathrm{A} \to \mathrm{B} \to \mathrm{C} \to \mathrm{D} \to \mathrm{A}$ in Fig. 1. The cycle consists of the following four processes: in the processes $\mathrm{A} \to \mathrm{B}$ and $\mathrm{C} \to \mathrm{D},$ $D_0$ is kept fixed at a large diffusivity, $D_L,$ and a small diffusivity, $D_S,$ respectively, whereas in the processes $\mathrm{B} \to \mathrm{C}$ and $\mathrm{D} \to \mathrm{A}$, $T$ remains constant at high temperature, $T_H$, and low temperature, $T_C$, respectively. The values of $c$ at A, B, C, and D are given by $c_1$, $c_2$, $c_3$, and $c_4$, respectively. Accordingly, the following equalities hold:

$$c_1 T_C = c_2 T_H = D_L, \tag{14}$$

$$c_3 T_H = c_4 T_C = D_S, \tag{15}$$

leading to the relation given by

$$\frac{c_1}{c_2} = \frac{c_4}{c_3} \quad (>1). \tag{16}$$

The analog of the quantity of heat is increased in the process $\mathrm{A} \to \mathrm{B}$, which is found to be

$$Q_D = \int_{(\mathrm{A})}^{(\mathrm{B})} d' Q_D = - \int_{c_1}^{c_2} dc \, \frac{D_L}{c} = D_L \ln \frac{c_1}{c_2}. \tag{17}$$

It is obvious from Eq. (7) that this is equal to the analog of work described by $W_{\mathrm{AB}}$ in the process. The analogs of work in other processes are also given by

$$W_{\mathrm{BC}} = \int_{(\mathrm{B})}^{(\mathrm{C})} d'W_D = -\int_{c_2}^{c_3} dc\, T_H \ = D_L - D_S, \tag{18}$$

$$W_{\mathrm{CD}} = \int_{(\mathrm{C})}^{(\mathrm{D})} d'W_D = -\int_{c_3}^{c_4} dc\, \frac{D_S}{c} \ = -D_S \ln \frac{c_4}{c_3}, \tag{19}$$

$$W_{\mathrm{DA}} = \int_{(\mathrm{D})}^{(\mathrm{A})} d'W_D = -\int_{c_4}^{c_1} dc\, T_C \ = D_S - D_L, \tag{20}$$

where Eqs. (14) and (15) have been used at the third equality in Eqs. (18) and (20). Therefore, the analog of the total work extracted in the cycle is calculated to be

$$\begin{aligned} W_D &= W_{\mathrm{AB}} + W_{\mathrm{BC}} + W_{\mathrm{CD}} + W_{\mathrm{DA}} \\ &= \left( 1 - \frac{D_S}{D_L} \right) Q_D, \end{aligned} \tag{21}$$

where Eq. (16) has been used at the second equality.

Consequently, we see that the efficiency of the present heat-like engine during the cycle, $\eta_D = W_D / Q_D$, is formally equivalent to the Carnot efficiency [45,46]:

$$\eta_D = 1 - \frac{D_S}{D_L}, \tag{22}$$

showing the dependence only on the quantity preserving constant in the process involving the increase/decrease of the analog of the quantity of heat, which is the average diffusivity, in conformity with the Carnot efficiency, in which the temperature preserves constant in the isothermal process involving the absorption/rejection of the quantity of heat. This result becomes more manifest if the case when $c_2 = c_4$ is considered, for which the efficiency in Eq. (22) is reduced to

$$\eta_D = 1 - \frac{T_C}{T_H}. \tag{23}$$

## 4. Total entropy change

In this section, we show that the total change of the entropy associated with diffusivity fluctuations vanishes during the cycle, which is a natural result since the exponential law in Eq. (2) is analogous to the canonical distribution. For it, we need to evaluate the entropy change given, in the continuum limit, by

$$\delta\tilde{S} = \frac{\delta' Q_D}{T} + S\,\delta c, \tag{24}$$

in which $S$ is determined up to the additive constant, leading to disappearance of the associated term during the cycle. The entropy changes in the four processes are given by

$$\tilde{S}_{\mathrm{AB}} = \int_{(\mathrm{A})}^{(\mathrm{B})} d\tilde{S} = \int_{c_1}^{c_2} dc \ln D_L = (c_2 - c_1)\ln D_L, \tag{25}$$

$$\tilde{S}_{\mathrm{BC}} = \int_{(\mathrm{B})}^{(\mathrm{C})} d\tilde{S} = \int_{c_2}^{c_3} dc\left[\,1 + \ln(cT_H)\,\right] = (c_3 - c_2)\ln T_H + c_3 \ln c_3 - c_2 \ln c_2, \tag{26}$$

$$\tilde{S}_{\rm CD} = \int_{\rm (C)}^{\rm (D)} d\tilde{S} = \int_{c_3}^{c_4} dc \ln D_S = ( c_4 - c_3 ) \ln D_S , \tag{27}$$

$$\tilde{S}_{\rm DA} = \int_{\rm (D)}^{\rm (A)} d\tilde{S} = \int_{c_4}^{c_1} dc \left[ 1 + \ln( c T_C ) \right] = ( c_1 - c_4 ) \ln T_C + c_1 \ln c_1 - c_4 \ln c_4 , \tag{28}$$

where Eqs. (7) and (8) have been used at the second equality in Eqs. (25) and (27). Clearly, from Eqs. (14) and (15), the total entropy change vanishes:

$$\begin{aligned} \tilde{S}_{\rm total} &= \tilde{S}_{\rm AB} + \tilde{S}_{\rm BC} + \tilde{S}_{\rm CD} + \tilde{S}_{\rm DA} \\ &= 0. \end{aligned} \tag{29}$$

**5. Effect of slowly varying fluctuation on efficiency**

Finally, we make a comment on the effect of the slow variation of diffusivity fluctuations on the efficiency in Eq. (22), which is not studied in the previous work in Ref. [33]. As mentioned in Sec. 2, the fluctuation distribution is supposed to vary slightly from the exponential law on a large time scale. Accordingly, a natural question may arise: how does the efficiency tend to be of the form in Eq. (22) as a time-dependent fluctuation distribution, $P(D, t)$, which is treated purposely in the continuous case, approaches the exponential law? To answer it, here we examine the approach of diffusing diffusivity in Ref. [13] widely discussed in the literature, which has originally been developed for the case of normal diffusion and has also been studied for the case of anomalous diffusion, e.g., in Refs. [11,36]. From Eq. (22), the dynamical behavior of the average diffusivity is of particular interest.

Dynamical evolution of the fluctuation distribution is described by the following advection-diffusion-type equation [13]:

$$\frac{\partial P(D,\, t)}{\partial t} = -\frac{\partial J(D,\, t)}{\partial D} \tag{30}$$

with $J(D,t)$ being the probability current defined by

$$J(D,\, t) = -k\,\frac{\partial P(D,\, t)}{\partial D} - s\, P(D,\, t), \tag{31}$$

where $k\ (>0)$ denotes the diffusivity of diffusivity, whereas $-s\ (<0)$ stands for the bias of the diffusion of diffusivity, provided that the reflecting boundary conditions, $J(0,t)=0$ and $\lim_{D \to \infty} J(D,\, t) = 0,$ are satisfied. It is understood that the distribution and its derivative in terms of the diffusivity vanish in the limit $D \to \infty.$ As shown in Ref. [13], the stationary solution of Eq. (30) becomes identical to the exponential law in Eq. (1) under the relation $D_0 = k/s.$ Therefore, the average diffusivity in Eq. (22) is chosen in such a way that it fulfills this relation.

Interestingly, in Ref. [36] (see also Ref. [37]), it has been proved, with the initial condition, $P(D,\, 0) = \delta(D),$ that the entropy production rate of diffusivity fluctuations, i.e., the time derivative of a time-dependent entropy based on $S,$ under Eq. (30) is manifestly positive, showing the monotonic increase of the entropy. Such an entropy is introduced as $S = -\int dD\, P(D,t) \ln P(D,t),$ the time derivative of which is calculated

to be

$$\frac{dS}{dt} = \frac{1}{k} \int dD \, \frac{\left[\, J(D,t) \,\right]^2}{P(D,t)} + \frac{s}{k} \frac{d\langle D \rangle_t}{dt} \tag{32}$$

with $\langle D \rangle_t$ being the average of the diffusivity with respect to $P(D, t)$, giving rise to $d\langle D \rangle_t / dt = kP(0, t) - s,$ provided that $\lim_{D\to\infty} P(D, t) = 0.$ The solution of Eq. (30) under the initial condition is then found to be given by [47]:

$$P(D, t) = \frac{1}{\sqrt{\pi k t}} \exp\left[ -\frac{(D + st)^2}{4kt} \right] + \frac{s}{k\sqrt{\pi}} \, e^{-\frac{sD}{k}} \int_{\frac{D - st}{2\sqrt{kt}}}^{\infty} dy \exp(-y^2), \tag{33}$$

with which it turns out that $\langle D \rangle_t$ monotonically increases in time:

$$\frac{d\langle D \rangle_t}{dt} = \sqrt{\frac{k}{\pi}} \int_{\sqrt{t}}^{\infty} dy \, y^{-2} \exp\left( -\frac{s^2 y^2}{4k} \right), \tag{34}$$

for which a slightly different form from that in Ref. [36] is used, showing from Eq. (32) that $dS / dt > 0.$

Recalling that we are considering the fluctuation distribution slightly different from the exponential law, a point is that time appearing in the integral corresponds to a large but finite one, meaning that Eq. (34) is less sensitive to $k$ and $s,$ except for the prefactor, which has the square-root dependence of $k.$ Thus, the tendency to reach the average diffusivity in time is dominantly determined by the diffusivity of diffusivity. It is interesting to see that the effect of slowly varying fluctuation on the efficiency is related to such a square-root nature. This observation indicates that it appears meaningful to study the efficiency from the viewpoint of diffusing diffusivity.

## 6. Conclusion

We have reported two recent topics on the formal thermodynamic analogy of intracellular diffusivity fluctuations in normal/anomalous diffusion. Based on the structural similarity of the exponential diffusivity fluctuation to the canonical distribution, the analogs of the internal energy, work, and the quantity of heat have been identified, and the Clausius-like inequality for the entropy relevant to the fluctuations has been established. To quantify extraction of the diffusivity change as the analog of work, the analog of the heat engine has also been constructed in a cycle constituted by the thermodynamic-like processes that are realized by compression/expansion of cells and the change of temperature. The efficiency of this heat-like engine has turned out to be formally consistent with the Carnot efficiency, together with the result that the total entropy change in the cycle disappears. We have also made a comment on the effect of slowly varying fluctuation on the efficiency.

## Acknowledgements

The present article is based on author’s talk at the 50th Conference of the Middle European Cooperation in Statistical Physics (25-29 March 2025, Dubrovnik, Croatia). He would like to thank the organizers of the conference for providing him with opportunity to give a talk. He was supported by a Grant-in-Aid for Scientific Research from the Japan

Society for the Promotion of Science (No. 21K03394).

**Figure 1**

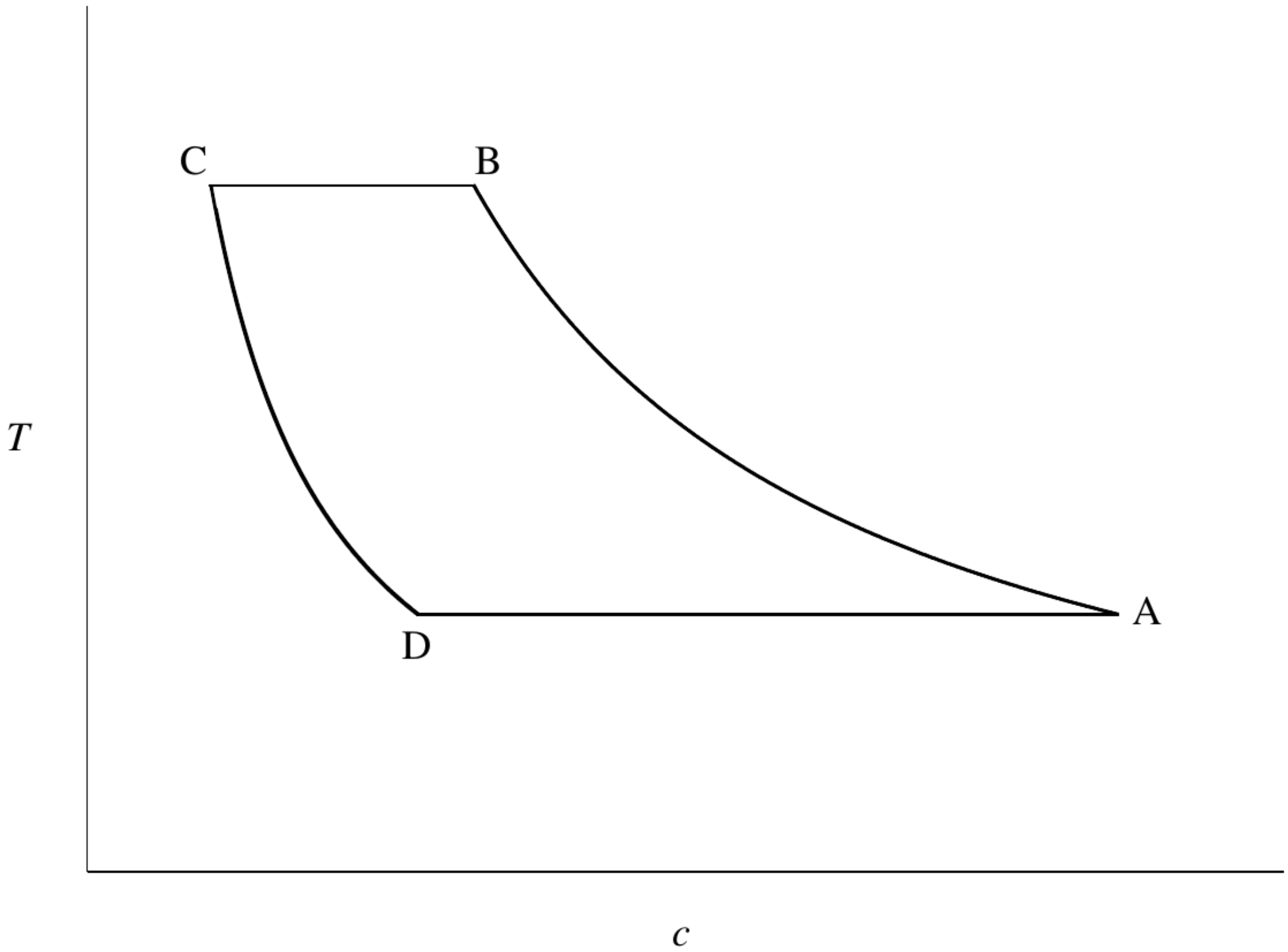

# Figure Caption

**Figure 1**

The Carnot-like cycle in the plane of the factor $c$ and average local temperature $T$. In the processes $\mathrm{A} \to \mathrm{B}$ and $\mathrm{C} \to \mathrm{D}$, the value of the average diffusivity $D_0$ is kept fixed, for which $T \propto 1/c$. In the processes $\mathrm{B} \to \mathrm{C}$ and $\mathrm{D} \to \mathrm{A}$, the value of $T$ remains constant.